\documentstyle[psfig,epsfig]{nature_wide}
\newcommand{\gtsim}{\mbox{{\raisebox{-0.4ex}{$\stackrel{>}{{\scriptstyle\sim}}
$}}}}
\newcommand{\ltsim}{\mbox{{\raisebox{-0.4ex}{$\stackrel{<}{{\scriptstyle\sim}}
$}}}}
\title{\bf The inevitable youthfulness of known high-redshift \\ 
radio galaxies}
\author{Katherine M.\ Blundell \& Steve Rawlings
\institute{\em University of Oxford, Astrophysics, Keble Road, Oxford, OX1 3RH}} 

\headertitle{The Youth--Redshift Degeneracy}
\mainauthor{Blundell \& Rawlings}

\summary{Some galaxies are very luminous in the radio part of the
spectrum. These `radio-galaxies' have extensive (hundreds of
kiloparsecs) lobes of emission powered by plasma jets originating
at a central black hole\cite{Beg84}.  Some radio-galaxies can be
seen at very high redshifts\cite{Raw96}, where in principle they
can serve as probes of the early evolution of the Universe.  Here
we show that for any model of radio-galaxy evolution in which the
luminosity decreases with time after an initial rapid increase
(that is, essentially all reasonable models\cite{Bal82}), all
observable high-redshift radio-galaxies must be seen when the
lobes are less than $10^7$ years old.  This means that
high-redshift radio galaxies can be used as a
high-time-resolution probe of evolution in the early Universe.
Moreover, this result explains many observed trends of
radio-galaxy properties with
redshift\cite{McC87,Kap87,Bar88,Blu99,Gar91,Hug98}, without
needing to invoke explanations based on cosmology\cite{Dal94} or
strong evolution of the surrounding intergalactic medium with
cosmic time\cite{Bar88}, thereby avoiding conflict with current
theories of structure formation\cite{Eke96}.  }

\dates{11 December 1998}{13 March 1999}

\begin{document}
\maketitle 
\noindent
The luminosity $P$ of the expanding lobes of a classical double radio
galaxy with age $t$ depends on the bulk kinetic power delivered by the
jets $Q$.  It will also be influenced by its environment, which we
parameterize at radius $r$ as $n(r) \propto r^{-\beta}$, where $n(r)$ is
the number density of gas particles.  A good empirical fit to radial
density profiles\cite{Rey96,Neu99} is obtained using $\beta = 1.5$.  It is
also a reasonable approximation to the so-called `universal' density
profile which emerges from $N$-body simulations of structure
formation\cite{Nav97}.  We normalise the density profile by consideration
of (i) detections of thermal X-ray haloes\cite{Cra96}, (ii) weak and
strong gravitational lensing\cite{Bow97,Del97}, (iii) galaxy
counting\cite{Hil91}, and (iv) studies of radio
depolarisation\cite{Gar91}, obtaining a characteristic number density
$n_{100}$ at radius 100\,kpc of $n_{100} = 2 \times 10^3\,{\rm m^{-3}}$.
Falle\cite{Fal91} found the length of a radio-galaxy $D$ grows as
\begin{equation}
\label{eq:falle}
D \propto \biggl(\displaystyle\frac{t^3 Q}{n_{100}}
              \biggr)^{\frac{1}{5-\beta}}.
\end{equation}
A simple model for a radio-source's luminosity evolution assumes that $Q$
remains constant as $t$ increases.  Combining equation~(\ref{eq:falle})
with a minimum-energy estimate of the energy stored in the lobes leads to:
\begin{equation}
\label{eq:lum}
P \propto Q^{{\ (26 - 7\beta)}/[4(5 - \beta)] }
             n_{100}^{9/[4(5 - \beta)]}t^{\ (8 - 7\beta)/[4(5 - \beta)]}.
\end{equation}
Using $\beta = 1.5$ gives
\begin{equation}
P \propto Q^{31/28} n_{100}^{9/14}t^{-5/28}.
\end{equation}
Equations~(\ref{eq:falle}) and (\ref{eq:lum}) can be used to
produce evolutionary tracks of $P$ versus $D$ for radio-sources
of given $Q$ and $n_{100}$ (Fig.\,1a).  A model\cite{KDA97} which
incorporates synchrotron cooling, adiabatic expansion losses and
inverse Compton scattering off the cosmic microwave background
modifies these tracks slightly (Fig.\,1a) and at large $D$
introduces downward curvature.

Our study of the properties and dependencies of complete
samples\cite{Blu99} of radio-sources led us to a model which
includes the r\^{o}le of the hotspots (the compact, high
surface-brightness regions at the end of jets) in {\em (i)}
governing the energy distribution of the particles injected into
the lobes and {\em (ii)} promoting increasing expansion losses
throughout a radio-galaxy's life.  In this model the $P$--$D$
tracks are, for a given environment, steeper (see Fig.\,1b) than
for the models in Fig.\,1a.  After a very rapid initial increase,
the luminosity of the radio lobes will {\sl decrease} as the
source grows, for $\beta$ \gtsim\ 0.3 in the case of the last
model (Fig.\,1b), and for $\beta > 8/7$ in the case of the first
and second models (described by equation (2) and ref 20
respectively) (Fig.\,1a).  With $\beta \sim 1.5$ radio luminosity
inevitably declines with $t$.

Application of a low-frequency flux-limit will inevitably lead to
more distant sources selected by the survey being more radio
luminous (this is called the Malmquist bias).  From equation
\ref{eq:lum}, this means that the most distant objects in a
survey are drawn from a higher-$Q$ population. Since the
dependence of $P$ on $Q$ is stronger than on $n_{100}$, and since
the dynamic range in $Q$ is larger than that of $n_{100}$,
jet-power $Q$ is the dominant factor, though density $n_{100}$
may be an important source of subtle biases.  If, as we have
argued, the luminosities of radio-sources decrease as they age,
more distant sources fall through the flux-limit sooner than
low-redshift sources.  Since our light-cone intercepts a
radio-source at a random point in its life-time\cite{Blu99}, it
is only the high-redshift radio-sources which are intercepted by
our light-cone when they are {\sl young} and luminous which can
be selected by these surveys.  {\em Any model involving a
declining luminosity-evolution will give a `youth-redshift
degeneracy'.}  Although it has been demonstrated that the
low-frequency surveys\cite{Ril89} are unlikely to be much
affected by surface-brightness dimming, any effect will be in the
sense to make old radio-galaxies at high-$z$ even harder to
detect.

The youth-redshift degeneracy is highly relevant to our
understanding of the alignment effect\cite{McC87}, the optical
and infra-red light which is aligned along radio-jet axes.  Where
this is caused by star-formation, it will be more easily
triggered close into the host galaxy or within the product of a
recent merger (assuming this is the jet-triggering
event\cite{San88}) than at distances further out sampled by the
head of an expanding radio-source later in its lifetime.  Where
this is caused by dust-scattered quasar light, the certain
youthfulness of distant radio-galaxies alleviates the near
discrepancy\cite{DeY98} between radio-source ages and the
time-scale for which dust grains can survive in the presence of
shocks caused by the advancing radio-jets.  The `youth--redshift
degeneracy' is consistent with the finding that the smallest
sources in a sample of $z \sim 1$ radio-galaxies (all with very
similar luminosities) are those which are most aligned with
optical emission\cite{Bes96}. Indeed, Best et al.\cite{Bes96}\
remarked that the sequence of changing optical aligned structure
with increasing radio size could be naturally interpreted by
comparing it with different phases of the interaction of the
radio jets with the interstellar and intergalactic medium as the
radio-sources age.

Barthel and Miley\cite{Bar88} had suggested that higher redshift
environments are denser and inhomogeneous than at low redshift since
they found increased distortion in the structures of their high-$z$ sample
of steep-spectrum quasars compared with their low-$z$ sample.  Sources
which are younger may have the passage of their jets considerably more
disrupted where there is a higher density and greater inhomogeneity in the
ambient post-recent-merger environment.  A general trend of denser
inter-galactic environments at high-$z$ cannot be inferred from their
result.

The mild linear size evolution observed in low-frequency selected
samples of classical double radio-sources\cite{Kap87,Bar88,Blu99}
arises because the high--$z$ sources are younger, hence tend to
be shorter.  It is the positive dependence on jet-power of the
rate at which the lobe-lengths grow (equation~\ref{eq:falle})
which contributes to the linear size evolution being as mild as
it is\cite{Blu99}.  Since we find it unavoidable that objects
found at high-redshift will be younger than those at low
redshift, the use of classical double radio sources as
`standardizable' rods\cite{Dal94} is beyond reach.  Fig.\,2
illustrates the difficulty of distinguishing between different
underlying cosmic geometries when more dramatic influences, such
as the youth-redshift degeneracy, and variations in source
environments, are at work.  Garrington \& Conway\cite{Gar91} have
found a tendency for depolarisation to be higher in sources with
a higher $P$ and/or $z$.  Objects with higher $P$ or $z$ which
are younger will be in much more recently merged environments
with the consequence that inhomogeneities in density or magnetic
field will more readily depolarise the synchrotron radiation from
the lobes, in addition to being closer in to the centre of the
potential well.

Many of the highest--$z$ radio-galaxies have gas masses
comparable to gas-rich spiral galaxies\cite{Hug98}, and inferred
star-formation rates which, in the local Universe, are rivalled
only by galaxy-galaxy mergers like Arp 220\cite{Gen98}.  If
high-$z$ objects are being viewed during a similar merging of
sub-components the associated star formation could be responsible
for a significant fraction of the stellar mass in the remnant
galaxy.  Since we have shown (Fig\,1) that the high--$z$
radio-galaxies --- those detected by SCUBA --- are necessarily
young (\ltsim\ $10^7$ years), and since the whole merger must
take a few dynamical crossing times, or $10^{8-9}$ years, the
implication is that the event which triggered the jet-producing
central engine is synchronised with massive star formation in a
gas-rich system, perhaps as material streams towards the minimum
of the gravitational potential well of the merging system.  The
youth-redshift degeneracy may explain why few lower-$z$
radio-galaxies show similarly large (rest-frame) far-infrared
luminosities compared to the high-$z$ population: they are being
observed significantly longer after the jet-triggering event.

The dramatic decrease in the co-moving space density\cite{Lon66}
of radio-galaxies from $z \sim 2$ to the local Universe is most
likely to be explained by the probability of suitable
jet-triggering events like galaxy-galaxy mergers\cite{San88}
being much lower now than at earlier cosmic epochs.  Cygnus A is
a local example demonstrating that this probability is not yet
zero.  Since efficacious interactions and mergers require slow
relative motions\cite{Aar80}, rich galaxy groups are excellent
sites for jet-triggering events by virtue of their high galaxy
number density but low velocity dispersion prior to
virialisation.  This is evinced by studies of the dynamics of
rich groups containing luminous quasars\cite{Ell91} which find
anomalously low velocity dispersions for systems of high galaxy
number density.  But if systems, whose galaxy number densities
are sufficiently high that galaxy encounters are frequent, have
formed into large virialised clusters by $z \sim 0$ then the
galaxies' encounter velocities will be prohibitively high.  This
explains why radio-source environments may be seen to get richer
between $z \sim 0$ and $z \sim 0.5$\cite{Pre88,Hil91}.  The space
density evolution of radio-sources may therefore be understood in
the context of current structure-formation theories\cite{Eke96}.
Only in systems where the merger of sub-groups is still on-going
are jets likely to be triggered in the local Universe: Cygnus A
was recently shown to be embedded in an ongoing merger of two
sub-clusters\cite{Owe97,Mar98}.

Fig.\,2 shows the location on the $P$--$D$ plane of the most
extreme redshift ($z > 3$) radio galaxies known.  Fig.\,2a
($\Omega_{\rm M} = 1$, $\Omega_{\Lambda} = 0$) ($\Omega_{\rm M}$
and $\Omega_\Lambda$ are the dimensionless density parameters for
matter and the cosmological constant respectively) suggests that
not only are the extreme redshift sources younger than the most
distant objects in the 3C sample\cite{Lai83} but that they have
similar jet-powers. Fig.\,2b ($\Omega_{\rm M} = 0.3$,
$\Omega_{\Lambda} = 0.7$) requires that objects with higher
jet-powers ($\sim 10^{40}$ W) exist beyond those detected in the
$z < 2$ Universe sampled by 3C.  The extreme values of $Q$
required in even the most conservative cosmological model
constrains their jet-producing AGNs to be only those with the
most massive black holes.  A jet-power of $10^{40}$\,W
corresponds to a 100\%-efficient Eddington-rate process for a
$10^9$ M$_\odot$ black hole. Unless one is prepared to postulate
significantly larger black hole masses, it seems difficult to
avoid the conclusion that by focussing on these objects, we are
selecting objects from the extreme end of the distribution
function of black-hole masses.  Interpretation of the properties
of distant radio galaxies must account for this effect alongside
the youth-redshift degeneracy.  We remark that the term
`Cosmological evolution' should not be used to describe trends
with redshift where they arise because of well-understood
physical mechanisms.  The youth-redshift degeneracy is one
example where trends with source age have become confused with
changes with cosmic epoch.  However, the youth-redshift
degeneracy brings two important benefits: first, since
extreme-redshift radio galaxies are young, all with similar $Q$,
they deliver the fine time-resolution required for the solution
of problems which it may be difficult to study with objects like
optically-selected quasars, whose ages are indeterminate:
examination of the environments of distant radio galaxies
provides a snapshot of the host galaxy evolutionary status just
after the jet-triggering event. Second, at redshift $\sim 4$ we
observe radio galaxies $\sim 1$ Gyr after the Big Bang and in
environments which saw a jet-triggering event a time-step no
greater than $10^7$ years prior to that.  This time-step is over
an order of magnitude smaller than the dynamical crossing time of
a massive galaxy, and 2 orders of magnitude smaller than the age
of the Universe at the epochs probed, giving fine time-resolution
essential to any study of triggering (and hence merging) rates at
early cosmic epochs.

\smallskip
\noindent {\small {\bf Acknowledgements.} K.M.B.\ thanks the Royal
Commission for the Exhibition of 1851 for a Research Fellowship. }

\medskip
\noindent {\small Correspondence should be addressed to K.M.B. \\
(e-mail: kmb@astro.ox.ac.uk).} 

\newpage
\begin{table*}
\centering
\includegraphics{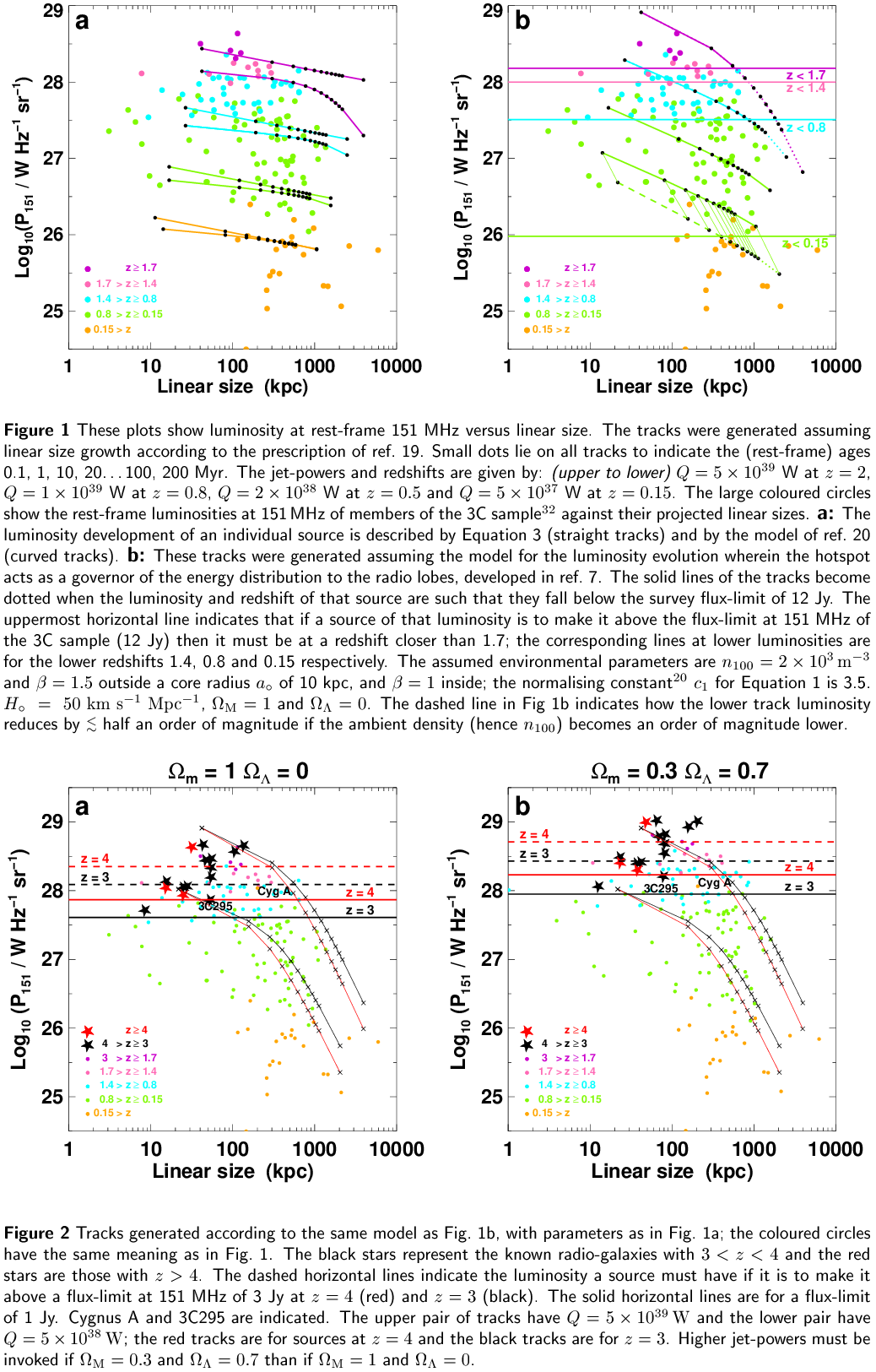}
\vbox to 200mm{\vfil}
\end{table*}
\end{document}